# Long and short range magnetism in the frustrated double perovskite $Ba_2MnWO_6$

Heather Mutch[1], Otto Mustonen[1], Helen C. Walker[2], Peter J. Baker[2], Gavin B. G. Stenning[2], Fiona C. Coomer,[3] Edmund J. Cussen[1*]

1. Department of Material Science and Engineering, University of Sheffield, Mappin Street, Sheffield, S1 3JD, United Kingdom

2. ISIS Pulsed Neutron and Muon Source, STFC Rutherford Appleton Laboratory, Harwell Campus, Didcot, OX11 0QX, United Kingdom

3 Johnson Matthey Battery Materials, Blount's Court, Sonning Common, Reading, RG4 9NH United Kingdom

* Corresponding author

The structural and magnetic properties of the face-centered cubic double perovskite $Ba_2MnWO_6$ were investigated using neutron powder diffraction, DC-magnetometry, muon spin relaxation and inelastic neutron scattering. $Ba_2MnWO_6$ undergoes Type II long-range antiferromagnetic ordering at a Néel temperature of 8(1) K with a frustration index, $f \approx 8$. Inelastic neutron scattering was used to identify the magnetic coupling constants $J_1$ and $J_2$, which were found to equal -0.080 meV and -0.076 meV respectively. This indicated that both of the magnetic coupling constants were antiferromagnetic with similar magnitudes, which is in contrast to other known $3d$ metal double perovskites $Ba_2MWO_6$. Above the Néel temperature, muon spin relaxation measurements and inelastic neutron scattering techniques identify a short-range correlated magnetic state that is similar to that observed in the archetypical face-centered cubic lattice antiferromagnet MnO.

## I. INTRODUCTION

Geometrical frustration is a phenomenon which can prevent antiferromagnetic systems from achieving a unique ground state. This is driven purely by the structure of the material and primarily affects lattices which consist of connecting triangular plaquettes, such as planar kagome lattices, pyrochlores and the face-centred cubic (*fcc*) structure [1]. Geometrical frustration can result in a number of novel states such as spin glass [2,3], spin ice [4,5] spin liquids [6,7] and other complex states such as the valence bond glass [8,9]. Frustration has also been shown to cause unexpected behaviour above the Néel temperature ($T_N$), where short range correlations can take place [10,11]. Many of these novel spin states are found when magnetic cations are arranged in the *fcc* lattice and we have undertaken an extensive study of these frustration effects.

Magnetism in *fcc* lattice compounds can often be described using a simple $J_1$-$J_2$ Heisenberg model:

$$\hat{H} = -J_1 \sum_{\langle ij \rangle} \mathbf{S}_i \cdot \mathbf{S}_j - J_2 \sum_{(ij)} \mathbf{S}_i \cdot \mathbf{S}_j \qquad (1)$$

where $J_1$ is the nearest neighbour (NN) interaction, $J_2$ is the next nearest neighbour (NNN) interaction, $S_i$ is spin at site $i$, and the sums are taken over all NN ($J_1$) and NNN ($J_2$) interactions respectively. Using the above equation, antiferromagnetic interactions are denoted using $J < 0$ whilst $J > 0$ for ferromagnetic interactions. Three distinct antiferromagnetically ordered ground states are possible depending on the relative strengths of the $J_1$ and $J_2$ interactions: Type I, Type II and Type III. Type II antiferromagnetic order is the most commonly observed magnetic structure in *fcc* lattices. The Type II structure occurs when $J_2$ is dominant and antiferromagnetic, and as a result all next nearest neighbour magnetic cations couple antiferromagnetically. In this instance, magnetic frustration arises from the weaker $J_1$ interaction as not all nearest neighbour magnetic cations are able to couple antiferromagnetically [12,13].

MnO is an archetypal Type II *fcc* antiferromagnet and has been shown to experience frustration effects. In 1949, Shull and Smart reported the detection of antiferromagnetic peaks at 80 K using neutron diffraction in this material. This was the first time this signature of antiferromagnetism was directly observed and MnO was correctly identified to possess the propagation vector of ***k*** = (1/2 1/2 1/2) corresponding to Type II order [14]. Since this was reported MnO has been keenly investigated and was initially reported to experience a structural phase transition from cubic to rhombohedral at its magnetic transition temperature of 118 K. This transition was stated to enable long range antiferromagnetic order to take place within MnO, where $Mn^{2+}$ ions are arranged in a parallel configuration in (111) planes and any perpendicular (111) planes are arranged in an antiparallel manner [15,16]. However, this is now known to be the result of a monoclinic distortion to space group $C2$, rather than a rhombohedral distortion [17]. More recently, MnO has also been reported to exhibit a short-range correlated state above its transition temperature, which is thought to be the result of magnetic frustration within the system [18].

As well as simple oxides such as MnO, double perovskites such as $Ba_2MnWO_6$ also host magnetic cations on an *fcc* lattice. Like MnO, $Ba_2MnWO_6$ has a magnetic structure composed of high spin $Mn^{2+}$ cations ($3d^5$, $S = 5/2$). $Ba_2MnWO_6$ crystallises in the space group $Fm\bar{3}m$, remaining cubic until 2 K, with no structural phase transitions reported. This was initially studied in 1976 by Khattak, Cox and Wang, who reported a Type II antiferromagnetic structure with propagation vector $\bm{k}$ = (1/2 1/2 1/2) and a magnetic transition temperature of approximately 9 K. This was based on neutron powder diffraction and DC-magnetometry measurements [19]. Further studies completed on $Ba_2MnWO_6$, correlated fairly well with this initial assessment, also indicating a type-II AFM structure [20]. However, the possibility of short-range magnetic correlations similar to those present in MnO has not been investigated nor are the magnetic interactions $J_1$ and $J_2$ known.

This study presents a comprehensive investigation of the magnetic behaviour observed in $Ba_2MnWO_6$ utilising neutron powder diffraction, DC-magnetometry, muon spin relaxation and inelastic neutron scattering. Using this range of techniques, $Ba_2MnWO_6$ we have characterised the Type II antiferromagnetic order below a transition temperature of 8(1) K. Above this transition temperature, a short range correlated state, similar to that described in MnO [18], was detected using muon spin relaxation in conjunction with inelastic neutron scattering. Inelastic neutron scattering shows that the magnetic interactions $J_1$ and $J_2$ are both antiferromagnetic with almost equal magnitude, with $J_1$ the slightly stronger coupling constant. This is an unexpected trait of $Ba_2MnWO_6$ which is not observed in other similar double perovskites and is interesting as $J_2$ interactions remain unfrustrated within the magnetic lattice.

## II. EXPERIMENTAL

A powder sample of $Ba_2MnWO_6$ was prepared via solid state synthesis using stoichiometric quantities of $BaCO_3$, $MnO_2$ and $WO_3$. The reagents were mixed with extensive grinding and calcined in air at 800 °C. The powder was then thoroughly ground, pressed into 13 mm diameter pellets and subsequently heated to 1250 °C in a reducing atmosphere of 5% $H_2$ in $N_2$ for approximately 96 hours. These grinding and heating steps were repeated out until phase purity was achieved. A Panalytical X'Pert Pro diffractometer with a Cu $K\alpha$ X-ray source was used to determine the phase purity of the samples.

Time of flight neutron powder diffraction experiments observed the low temperature nuclear structure and the long-range magnetic structure. This was completed on GEM, the General Materials diffractometer at the ISIS Neutron and Muon Source, Didcot, Oxfordshire, UK [21]. Approximately 6 g of powder was placed in an 8 mm cylindrical vanadium can and the data were corrected for absorption. The collected data set can be found at the following reference [22]. Structural refinements against these data used the General Structure Analysis System (GSAS), and a convolution of the Ikeda-Carpenter and pseudo-Voigt functions to describe the peak shapes [23]. The background was modelled using a shifted Chebyshev function. The magnetic form factor of the $Mn^{2+}$ ions was taken from the neutron data booklet [24].

Magnetic susceptibility measurements were carried out using a Quantum Design MPMS3 based at the Materials Characterisation Laboratory at the ISIS Neutron and Muon Source. Temperature dependent magnetic susceptibility was measured between 2 K and 300 K under zero field cool (ZFC) and field cool (FC) conditions in an applied field of 1000 G. Approximately 100 mg of $Ba_2MnWO_6$ was placed in a gelatine capsule and held tightly in place using a small quantity of PTFE tape. The filled capsule was then inserted into a plastic straw for measurement.

Muon spin relaxation measurements were carried out on EMU at the ISIS Neutron and Muon Source, where zero, transverse and longitudinal fields were applied to the sample [25]. Around 5 g of $Ba_2MnWO_6$ was densely packed into silver foil and placed on a silver sample holder. Zero-field measurements were completed between 2 K and 30 K. A transverse field of 100 G was applied between 2 K and 30 K to observe the transition temperature. Longitudinal field measurements were carried out to study the decoupling of muon spins from the internal magnetic field. The measured data can be observed using the referenced DOI [26]. The data was analysed using the Mantid software [27].

Inelastic neutron scattering measurements were performed on the LET time-of-flight direct geometry spectrometer at the ISIS facility of the Rutherford Appleton Laboratory [28]. The sample was contained in a cylindrical aluminium can of diameter 10 mm and height 30 mm, which also contained helium exchange gas. The sample can was cooled by a helium cryostat. Data were collected using neutrons with incident energies of 3.7 meV at temperatures between 2 and 100 K for ~3h each, these measured data sets can be found using the following reference [29]. The data were reduced using the MantidPlot software package [27]. The raw data were corrected for detector efficiency and time independent background following standard procedures [30].

## III. RESULTS

### A. Diffraction

Laboratory X-ray diffraction data indicate that the single phase double perovskite $Ba_2MnWO_6$ had been synthesised, with the *fcc* structure arising from $Mn^{2+}/W^{6+}$ cation ordering over the octahedral sites of the perovskite structure as previously described [19,20].

Neutron powder diffraction was utilised to determine both the low temperature nuclear structure and to analyse the magnetic structure of $Ba_2MnWO_6$. This was carried out between temperatures of 2 K and 69 K. Throughout the temperature range there was no peak broadening or splitting and the nuclear phase could be accurately fitted using the *fcc* space group, $Fm\bar{3}m$, indicating no phase transition in the crystal structure down to 2 K. Conventional thermal expansion properties were exhibited by the material however a change in the slope of the lattice parameter could be observed around 9(1) K which is indicative of magnetostrictive behaviour [31]. A representative fitted diffraction pattern, collected at 2 K, and the thermal evaluation of the lattice parameters are presented in Figure 1.

In data sets collected at $T \leq 9$ K, a number of additional Bragg peaks were evident at higher d-spacings, whilst the data at shorter d-spacings remain unchanged, indicative of the onset long-range antiferromagnetic order. These peaks were readily indexed using a doubling of the crystallographic cell along each axis corresponding to a magnetic propagation vector $\boldsymbol{k} = (1/2\ 1/2\ 1/2)$.

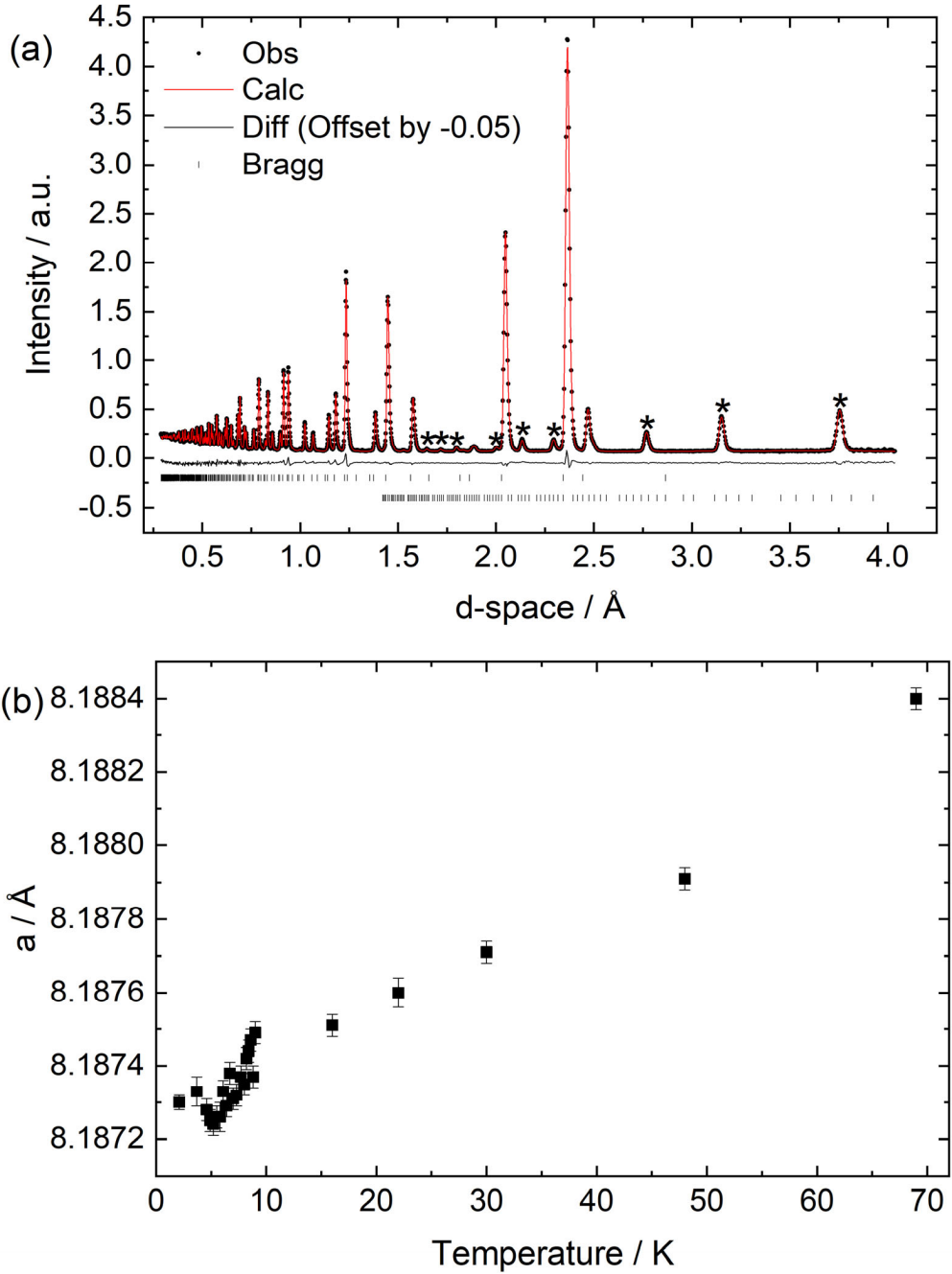

Fig. 1. (a) Neutron powder diffraction pattern of Ba$_2$MnWO$_6$ at 2 K. This contains both the nuclear phase (upper Bragg reflections) and the magnetic phase (lower Bragg reflections). The nuclear phase is characterised using the cubic space group $Fm\bar{3}m$, with lattice parameter $a$ = 8.18730(2) Å. The magnetic phase has the propagation vector $\boldsymbol{k}$ = (1/2 1/2 1/2). The solely magnetic peaks are outlined by an asterisk. (b) Thermal evolution of lattice parameter. A general increase is observed in the lattice parameter as the temperature increases, showing conventional thermal expansion behaviour.

As the direction of magnetic moments in cubic materials like $Ba_2MnWO_6$ cannot be determined precisely using powder diffraction, different potential magnetic structures were used to identify which best described the magnetic structure of $Ba_2MnWO_6$ [32]. This was completed using the model building in GSAS with a simultaneous checking approach using the FULLPROF code [33,34]. The neutron data sets were initially fitted using the established Ising spin model as outlined for $Ba_2MnMoO_6$, [35] followed by the non-collinear model described for $Ba_2Nd_{0.10}Mn_{0.90}MoO_6$, [36] but neither approach provided a good match to the data. Subsequently the magnetic structure of MnO was trialled. Despite the structural distortion observed in MnO below 118 K, this compound was considered to be a likely candidate due to the pseudo-*fcc* nature of the lattice and the presence of high spin $Mn^{2+}$ cations. In this model, the $Mn^{2+}$ magnetic moments within (111) planes are aligned in a parallel configuration but are aligned antiparallel to this in adjacent (111) planes. The direction of the magnetic moments alternate in parallel and antiparallel fashion along the $(11\bar{2})$ axes [18]. Rietveld refinements were completed using each of these models, which were identical bar the magnetic structure utilised. From this analysis the magnetic structure of MnO, gave the best fit to the observed data with $R_{wp} = 4.52$ and $\chi^2 = 4.73$, the output from this refinement is shown in Table I. This refinement yielded an effective magnetic moment of 4.5(1) $\mu_B$, the thermal evolution of which can be observed in Figure 2(a).

Table I. Structural parameters derived from Rietveld refinement against neutron powder diffraction collected on $Ba_2MnWO_6$ at 2 K. Bond valence sums (BVS) were calculated using tabulated values [44].

| Atom | Ba | Mn | W | O |
|---|---|---|---|---|
| $100 \times U_{iso}$ / Å$^2$ | 0.16(1) | 0.26(2) | 0.08(1) | 0.392(6) |
| $M - O$ / Å | 2.89732(1) | 2.1711(3) | 1.9225(3) | - |
| BVS | 2.32 | 2.14 | 5.98 | |

Space group $Fm\bar{3}m$; a = 8.18730(2) Å: Ba on ¼, ¼, ¼ ; Mn on 0, 0, 0; W on ½, 0, 0; O on 0.26519(3), 0 0. Ordered magnetic moment on $Mn^{2+}$ of 4.5(1) $\mu_B$

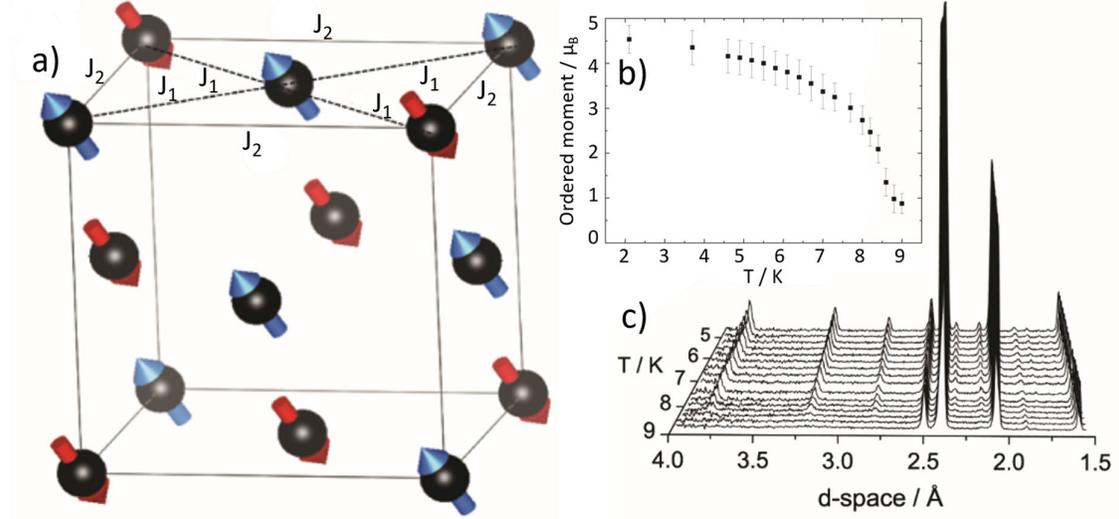

Fig. 2. (a) Schematic of the magnetic structure of $Ba_2MnWO_6$. Only magnetic $Mn^{2+}$ ions are depicted within the crystallographic structure, with antiparallel magnetic moments outlined in red or blue. The nearest neighbour (NN) interactions are depicted by the dashed line denoted by $J_1$, whilst the next-nearest neighbour (NNN) interactions are signified by $J_2$. (b) The thermal evolution of the ordered magnetic moment of $Ba_2MnWO_6$. This reaches a maximum value of 4.5(1) $\mu_B$ at 2 K, and decreases as the temperature tends to $T_N$. (c) Thermal evolution of the magnetic Bragg peaks from neutron diffraction data collected on GEM. The magnetic Bragg peaks are absent at 9 K and increase in magnitude as the temperature decreases.

## B. Magnetic properties

DC-magnetometry data collected on $Ba_2MnWO_6$ indicate an antiferromagnetic transition at approximately 8 K with no divergence between the FC and ZFC data shown in Figure 3. Curie-Weiss fitting of the data between 100 and 300 K yielded Curie and Weiss constants of $C = 4.91(1)$ cm$^3$ K mol$^{-1}$ and $\theta = -63(1)$ K. The Curie constant was used to calculate the effective magnetic moment, $\mu_{eff} = 6.3(3)$ $\mu_B$ which is close to that expected for the spin only value of high spin $Mn^{2+}$ where $\mu_{SO} = 5.9$ $\mu_B$. The frustration index (f), derived from the Curie-Weiss fit, $f = \theta_{CW}/T_N = 8$ indicates that $Ba_2MnWO_6$ is moderately frustrated, with a value below that of $f \geq 10$ that is considered to indicate a strongly frustrated system [1].

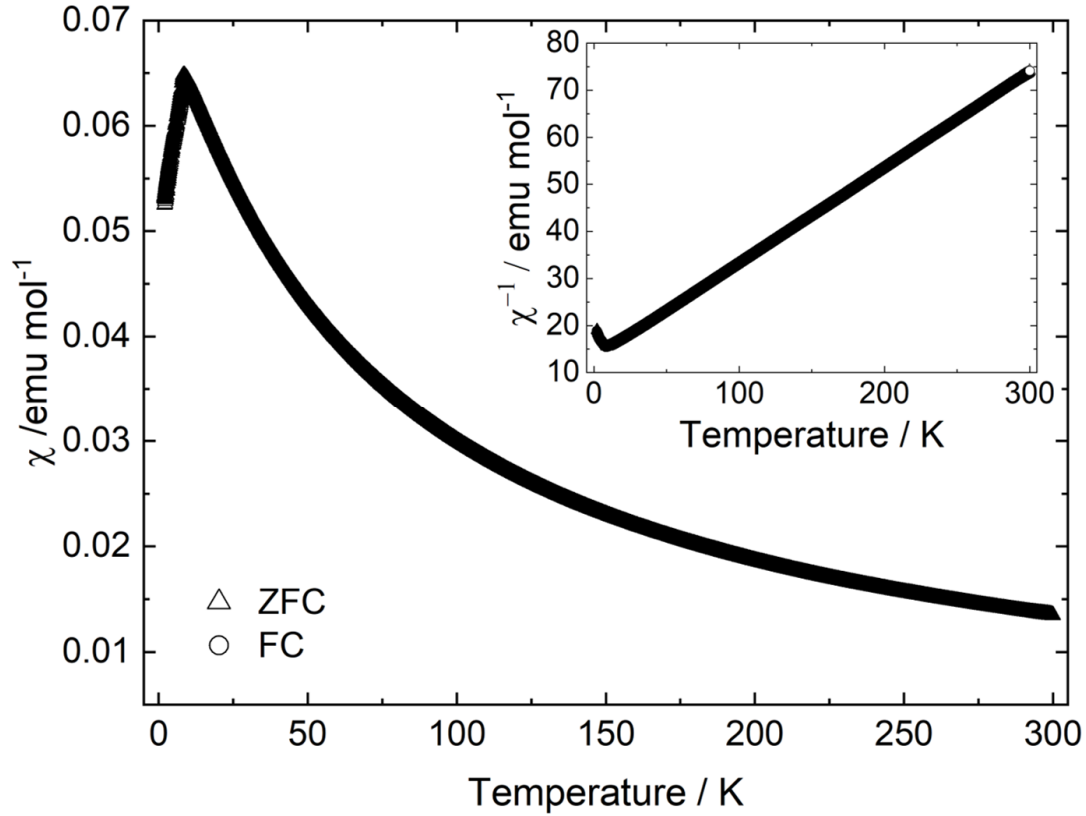

Fig. 3. DC magnetic susceptibility of $Ba_2MnWO_6$ measured in an applied field of 1000 G between 2 K and 300 K. The ZFC and FC susceptibility ($\chi$) are plotted on the main graph, whilst the inverse susceptibility ($\chi^{-1}$) can be seen in the inset. No divergence is observed between the ZFC and FC responses. The susceptibility curve indicates that $Ba_2MnWO_6$ goes through a transition to a long-range antiferromagnetic state at 8(1) K.

### C. Muon spin rotation and relaxation

The local nature of the muon spin relaxation measurements on $Ba_2MnWO_6$ was used to complement the bulk techniques outlined previously. Initially, an external transverse field (TF-µSR) of 100 G was applied to the system and measurements were carried out between 1.6 K and 70 K to identify the transition temperature accurately. When the sample is in a paramagnetic state, the muon spins couple to the applied transverse field resulting in a large oscillating signal as depicted in Figure 4(a). In the magnetically ordered state, the static internal fields are much stronger than the weak applied transverse field and the oscillation is no longer observed. Thus, the asymmetry of the oscillating signal in TF-µSR can be used to determine the transition temperature. Furthermore, the paramagnetic fraction of the sample at a given temperature can be estimated from the normalised magnitude of the oscillatory term relative to that observed in the paramagnetic regime at a high temperature of 70 K.

All of these data sets could be fitted using the sum of an oscillating decay, an exponential decay and a flat background. This is described by the following function:

$$A(t) = A_{osc}e^{-\lambda_{osc}t}\cos(2\pi ft + \phi) + A_2 e^{-\lambda_2 t} + A_3 \qquad (2)$$

where $A_{osc}$, $A_2$ and $A_3$ correspond to the asymmetries of the oscillating decay, the exponential decay and the constant signal respectively, $\lambda$ corresponds to the relaxation rate, $f$ is the frequency of the oscillation (which is dependent on the applied transverse field) and $\varphi$ is the phase of this oscillation. The decaying functions were related to any possible relaxation within the system, arising due to unpaired electrons whilst the flat background can be attributed to large static magnetic fields along the muon's initial spin direction.

Figure 4(c), shows the thermal evolution of the oscillating signal fraction in $Ba_2MnWO_6$. At 2 K, no asymmetry should be observed as the internal magnetic field is significantly greater than the applied transverse field. However, a small oscillation is observed which is attributed to muons implanted in the silver sample holder. As the temperature increases the fraction remains small until 8.25(8) K where the relaxing proportion of the material begins to increase, therefore this was identified as the transition temperature of the material. A further increase in the temperature shows a gradual increase in this relaxing proportion, and this was attributed to potential short-range magnetic correlations within the material. This gradual increase did not reach a plateau over the measured range up to 30 K.

Zero-field measurements were carried out between 1.6 K and 30 K. These showed characteristic low asymmetry below the ordering temperature which was found to increase with temperature. Between 1.6 K and 20 K the ZF-data sets could be fitted using:

$$A(t) = A_1 e^{-\lambda_1 t} + A_2 e^{-\lambda_2 t} + A_0 \qquad (3)$$

This again corresponds to both relaxing and static behaviour present within $Ba_2MnWO_6$. This leads to two separate relaxations, where $A_1$ corresponds to a slow relaxation and $A_2$ corresponds to a faster relaxation, with relaxation rates of $\lambda_1$ and $\lambda_2$ respectively. The persistence of relaxation above the transition suggests that short-range correlations exist within the system. At 30 K, the function reduces to:

$$A(t) = A_1 e^{-\lambda_1 t} + A_0 \qquad (4)$$

At which point the short correlations are no longer detected by the implanted muons.

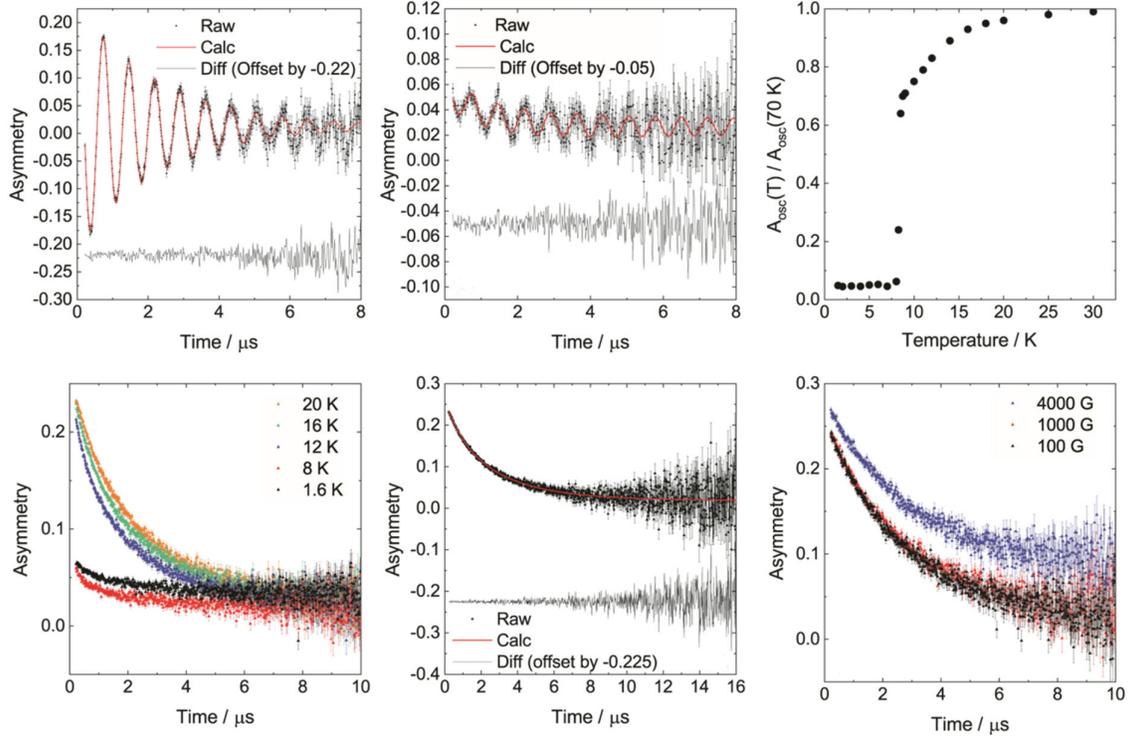

Fig. 4. (a) (b) TF- μSR measurement carried out at 20 K and 1.5 K respectively. At 20 K, oscillations are observed as muons couple to the perpendicular field, indicating that the muons observe paramagnetic behaviour. At 1.5 K, Ba$_2$MnWO$_6$, is magnetically ordered and therefore these oscillations are reduced in amplitude as only muons stopping in the silver sample holder do not experience the large magnetic fields inside the sample. (c) The asymmetry of the oscillating relaxation was then normalised and has been plotted against temperature. It is initially low between 2 and ~8 K before a sharp rise at the transition temperature, this is followed by a more gradual increase between 8.5 K and 30 K, which is attributed to short range correlations. (d) ZF measurements carried out at a range of temperatures, above the transition the relaxation has a high asymmetry, whilst below $T_N$ the initial asymmetry is damped. (e) ZF-measurement carried out at 20 K, fit using F = (A$_1$exp(-λ$_1$x)) + (A$_2$exp(-λ$_2$x)) + A$_3$. This fit is indicative of potential short range correlations as two separate relaxation rates are able to persist above this transition temperature. (f) The longitudinal field data sets collected at 30 K, this is significantly above the transition temperature and from this it is clear that any magnetic behaviour is not decoupled by the external field, not even at 4000 G. This could suggest short range correlations between the Mn$^{2+}$ ions.

Longitudinal field measurements carried out above the transition temperature also support the presence of short-range magnetic correlations. Above $T_N$, a full decoupling of the relaxation was not achieved when a longitudinal field of 4000 G was applied to the sample, as shown in Figure 4(f). If no magnetic interactions were taking place within $Ba_2MnWO_6$, then a full decoupling of the relaxation curve would be expected.

### D. Inelastic neutron scattering

Inelastic neutron scattering measurements were carried out at a range of temperatures both below and above the transition temperature. Below $T_N$ these could be used to quantify the magnetic coupling constants within the system. Above $T_N$, these could be used to give firm evidence of short range excitations taking place within $Ba_2MnWO_6$.

Initially, inelastic neutron scattering was carried out at 2 K, where distinct spin waves were observed within the system. This pattern was then compared to a SpinW simulation, based on the Hamiltonian presented in equation (1), to extract the values of the coupling constant using linear spin wave theory [37]. The spin wave spectra were well described by the $J_1$-$J_2$ Heisenberg model with NN $J_1$ and NNN $J_2$ interactions. From this practice it was found that $J_1$ = -0.080 meV, whilst $J_2$ = -0.076 meV. Figure 5 shows the level of agreement between the measured and simulated inelastic neutron scattering data. This results in a $J_2/J_1$ ratio of 0.95, indicating that both are of similar strength. Despite $J_1$ being stronger than $J_2$, this $J_2/J_1$ ratio corresponds to Type II magnetic order in the *fcc* $J_1$-$J_2$ model [12]. This is because in the *fcc* model all antiferromagnetic $J_2$ interactions can be satisfied simultaneously but that is not the case for $J_1$ interactions. When testing this model, a dominant $J_2$ was also investigated using $J_1$ = -0.010 meV and $J_2$ = -0.146 meV, however, it was clear that this did not correlate with what was observed within the system. This is most clearly demonstrated by taking vertical cuts through the data as a function of energy transfer, integrating between $|Q|$=0.9 and 1.1 Å$^{-1}$, and comparing these with cuts through the SpinW simulations for the two different models. As depicted in Figure 5(c), the model with a dominant $J_2$ is unable to reproduce the double peak structure seen in the data and the model with $J_2/J_1$=0.95.

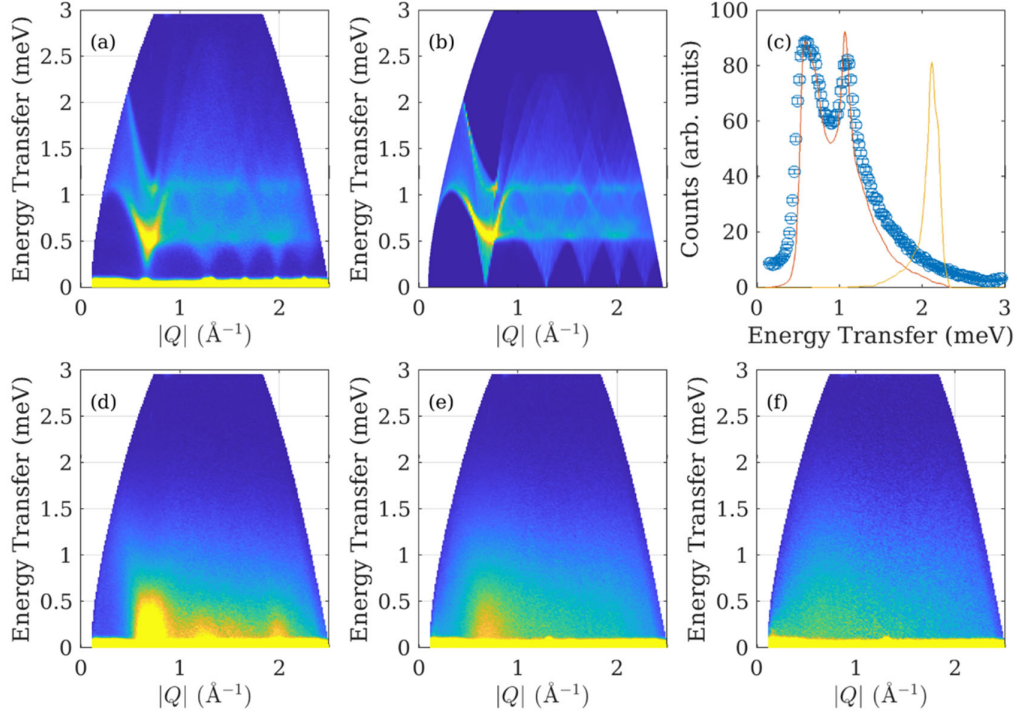

Fig. 5. (a) Inelastic neutron scattering data collected on LET at 2 K. Distinct spin waves can be observed due to long range order within the system. (b) A graphical output of the Spin W simulation carried out to obtain the coupling constants $J_1$ and $J_2$, these were determined to be -0.080 meV and -0.076 meV respectively, indicating antiferromagnetic interactions for both the NN and NNN neighbours. Both the experimentally observed data and the simulation tally to a reasonable degree indicating that the Heisenberg model used was consistent with the data. (c) A vertical cut through the data set for $|Q|= 0.9 - 1.1$ Å$^{-1}$, the left hand solid line shows the simulation output using $J_1 = -0.080$ meV and $J_2 = -0.076$ meV, where it is clear that the peaks in the data intensity are well reproduced, whilst the right hand solid line depicts only one peak in intensity which would be observed if $J_1 = -0.010$ meV and $J_2 = -0.146$ meV. It is clear from this that it is necessary for $J_1$ and $J_2$ to have similar magnitudes, where the peak splitting determines the ratio between $J_1$ and $J_2$. (d) (e) (f), the observed inelastic neutron scattering data collected at 12 K, 40 K and 100 K respectively. These are indicative of short range magnetic excitations being present within the system above $T_N$, which diminish as the temperature is increased.

Further inelastic neutron scattering measurements were carried out above the transition temperature at 12 K, 20 K, 40 K, as shown in Figures 5. At each of these temperatures distinct excitations could be observed which were established to be magnetic in nature as they disappeared at high temperatures and had greatest intensity at low momentum transfer. These magnetic excitations were found to originate from the same Q positions at the elastic line as the spin waves observed from the ordered state in the 2 K measurement. This suggests the short-range correlations are related to the magnetic order below $T_N$. These short range magnetic correlations can still be observed at 40 K, approximately five times the Neel temperature. At 100 K, these excitations could no longer be observed indicating that short range correlations had been disrupted by $k_B T$ at this temperature.

## IV. DISCUSSION

$Ba_2MnWO_6$ is a frustrated *fcc* antiferromagnet with a frustration index of $f = 8$. Each of the techniques outlined within this article show that the transition to a fully ordered antiferromagnet occurs at 8(1) K. A combination of neutron powder diffraction and inelastic neutron scattering show that $Ba_2MnWO_6$ displays characteristic Type-II antiferromagnetism where $J_1$ and $J_2$ have very similar magnitudes.

Both $J_1$ and $J_2$ interactions occur along Mn-O-W-O-Mn extended superexchange pathways with different angles. For the NN $J_1$ interaction this is a 90° pathway, whereas the NNN $J_2$ interaction occurs along lattice parameter $a$ in a 180° extended superexchange pathway. In the Type II magnetic structure, the $J_2$ interactions are unfrustrated as all NNN spins couple antiferromagnetically. The magnetic frustration arises through the antiferromagnetic $J_1$ interactions, as the *fcc* structure does not allow all NN spins to align antiferromagnetically. Since $J_1$ is stronger than $J_2$, $Ba_2MnWO_6$ is relatively frustrated with $f = 8$.

The strong $J_1$ observed in $Ba_2MnWO_6$ is unexpected and has not been reported in other $(Sr,Ba)_2MWO_6$ double perovskites where $M$ is a 3d transition metal. $Ba_2CuWO_6$ and $Sr_2CuWO_6$ are both Type II antiferromagnets which have been reported to have a dominant $J_2$ which is roughly 8-10 times greater than $J_1$. [38,39] Furthermore, recent theoretical studies into $Sr_2NiWO_6$, another Type II antiferromagnet, have indicated that $J_2$ is approximately 20 times larger than $J_1$, whilst previous experimental results suggest that $J_2/J_1$ is much higher than this at approximately 90 [40,41]. $Ba_2FeWO_6$ and $Sr_2FeWO_6$ have also been reported to adopt the Type II AFM structure, although the magnitudes of $J_1$ and $J_2$ in these materials have not been reported [42]. These systems cannot be directly compared to $Ba_2MnWO_6$ which remains undistorted and cubic as these both experience a tetragonal distortion within the space group $I4/m$. $Ba_2MnWO_6$ is also a $S = 5/2$ system compared to the $S = 1/2$ cuprate systems and $S = 1$ $Sr_2NiWO_6$. The difference in spin values and structure could result in these differences in behaviour. Unfortunately, similar Type II-AFM undistorted materials which retain their cubic nature down to 2 K such as $Ba_2CoWO_6$ [32] often do not have reported values for $J_1$ and $J_2$. However, $Ba_2NiWO_6$ is a cubic example of a $(Sr,Ba)_2MWO_6$ double perovskite which has a $J_2/J_1$ ratio of roughly 80, which is far greater than the 0.95 observed in $Ba_2MnWO_6$ [41,43]. Again, this stark contrast may be a result of the difference between the spin values of each system, however, it shows how unusual $Ba_2MnWO_6$ is. Despite the differences in the magnetic ion involved in these systems, the spin value and the structure each of these materials tend to adopt the Type II AFM structure which indicates how robust this magnetic ordering type is to the observed large variations in the strength of magnetic interactions.

Muon spin relaxation carried out in a transverse and longitudinal fields support the presence of a short-range correlated magnetic state above $T_N$ in $Ba_2MnWO_6$. Inelastic neutron scattering experiments confirmed the existence of this short-range correlated state, and suggested it is related to the magnetic order below $T_N$. While the precise nature of this state is unknown at this time, it could be related to the well-known correlated states in the similar *fcc* antiferromagnet MnO.

A number of parallels may be drawn between $Ba_2MnWO_6$ and MnO. Both materials possess $Mn^{2+}$ ions, and therefore a $3d^5$ electronic configuration corresponding to $S = 5/2$. These are present in a high spin state and therefore experience no orbital contribution to their magnetic behaviour. Both $Ba_2MnWO_6$ and MnO have Type II antiferromagnetic order with $T_N = 8(1)$ K and 160 K, respectively. The materials are relatively frustrated with frustration indexes $f = 8$ for $Ba_2MnWO_6$ and $f = 5$ for MnO. Similar values have been reported for the $J_2/J_1$ ratio of both materials (~1), although the magnitude of $J_1$ is smaller than $J_2$ in MnO, whilst the opposite is true in $Ba_2MnWO_6$. A lower $J_2/J_1$ ratio corresponds to a greater degree of frustration in the Type II magnetic structure, which explains the higher frustration index experienced by $Ba_2MnWO_6$ compared to MnO. MnO has also recently been reported to experience short range magnetic correlations above its transition temperature which was evidenced by diffuse neutron scattering alongside reverse Monte-Carlo simulations. [18] One large difference between the two materials is that MnO experiences a combined magnetic and structural transition at $T_N$, whilst $Ba_2MnWO_6$ remains cubic down to 2 K with no structural transitions. This suggests that magnetoelastic coupling is weaker in $Ba_2MnWO_6$.

## V. CONCLUSIONS

Neutron powder diffraction, DC-magnetometry, muon spin relaxation spectroscopy and inelastic neutron scattering were used to investigate the magnetic nature of $Ba_2MnWO_6$. Type-II long range magnetic order was observed by all of these techniques, which indicated a transition temperature of 8(1) K. The magnetic structure was characterised and found to have propagation vector, $\boldsymbol{k}$ = (1/2 1/2 1/2), although the direction of the magnetic moments cannot be accurately determined by powder diffraction, the magnetic structure was found to be best described by one related to another *fcc* lattice, MnO. Inelastic neutron scattering showed that $Ba_2MnWO_6$ is well described by the *fcc* $J_1$-$J_2$ Heisenberg model with magnetic coupling constants $J_1$ = -0.080 meV and $J_2$ = -0.076 meV. The similar magnitude and dominant $J_1$ is not observed in other related double perovskites making this an interesting case for comparison and may be caused by a combination of the cubic nature of $Ba_2MnWO_6$ and the large spin state of $S$ = 5/2. Muon spin relaxation measurements indicated that short range magnetic correlations take place above $T_N$. Inelastic neutron scattering measurements were able to confirm this, with magnetic excitations taking place at temperatures up to 40 K, close to five times the magnetic transition temperature.

## ACKNOWLEDGEMENTS

The authors acknowledge the financial support provided by the Leverhulme Trust Research Project Grant RPG-2017-109 and the Universities of Strathclyde and Sheffield. We also acknowledge and are grateful for the support given by Dr Ivan da Silva based at the ISIS Neutron and Muon Source with the powder neutron diffraction experiments. The ISIS Neutron and Muon Source and the Materials Characterisation Laboratory were both used extensively throughout this research and the authors are grateful for both being given access and the continuing support of the individuals within these facilities.